# Hyperbolic enhancement of photocurrent patterns in minimally twisted bilayer graphene

S. S. Sunku[1,2,*], D. Halbertal[1,*,Ɨ], T. Stauber[3], S. Chen[1], A. S. McLeod[1], A. Rikhter[4], M. E. Berkowitz[1], C. F. B. Lo[1], D. E. Gonzalez-Acevedo[1], J. C. Hone[5], C. R. Dean[1], M. M. Fogler[4], D. N. Basov[1]

[1] Department of Physics, Columbia University, New York, NY

[2] Department of Applied Physics and Applied Mathematics, Columbia University, New York, NY

[3] ICMM at CSIC, Madrid, Spain

[4] Department of Physics, University of California, San Diego, La Jolla, CA

[5] Department of Mechanical Engineering, Columbia University, New York, NY

* These authors contributed equally

Ɨ dh2917@columbia.edu (D.H.)

## Abstract

**Quasi-periodic moiré patterns and their effect on electronic properties of twisted bilayer graphene have been intensely studied. At small twist angle $\theta$, due to atomic reconstruction, the moiré superlattice morphs into a network of narrow domain walls separating micron-scale AB and BA stacking regions. We use scanning probe photocurrent imaging to resolve nanoscale variations of the Seebeck coefficient occurring at these domain walls. The observed features become enhanced in a range of mid-infrared frequencies where the hexagonal boron nitride substrate is optically hyperbolic. Our results illustrate the capabilities of the nano-photocurrent technique for probing nanoscale electronic inhomogeneities in two-dimensional materials.**

## Introduction

Twisted bilayer graphene (TBG), consisting of two graphene sheets rotated with respect to each other, has emerged as a tunable platform for studying exotic electronic phases. Transport experiments have revealed that when the graphene layers are twisted by a magic angle of $\theta \sim 1.1^\circ$, TBG can become a superconductor (*1*), a correlated insulator (*2*), or a quantum anomalous Hall insulator (*3*–*5*). A key feature of TBG is the moiré superlattice: a long-range variation in the atomic stacking arising from geometric interference of the lattice periodicities in the two graphene sheets. Scanning probe studies of TBG with $\theta \sim 1.1^\circ$demonstrated spatial variations in the electronic properties occurring on the length scale of tens of nanometers (*6*–*9*).

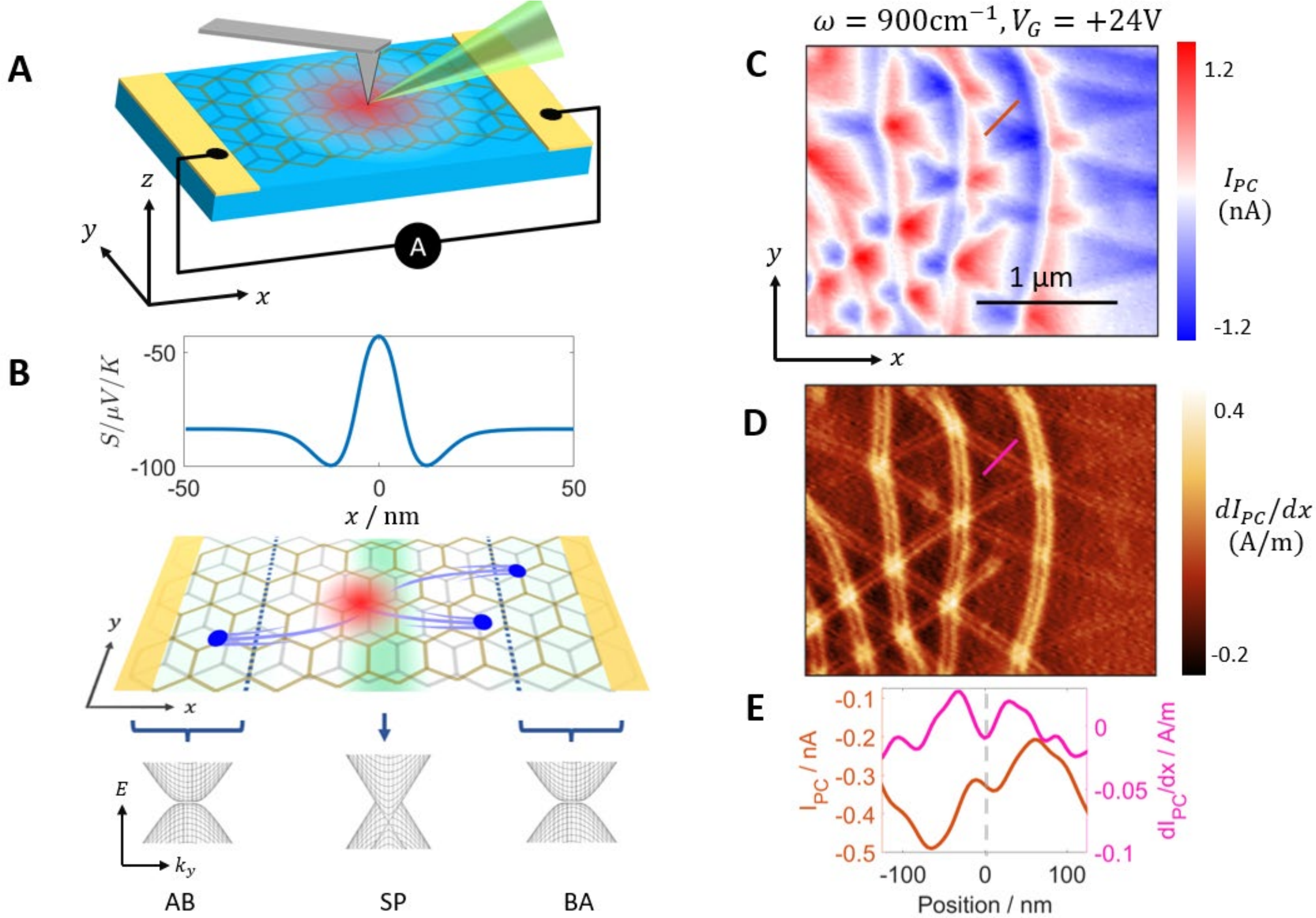

**Figure 1 | Photocurrent in minimally twisted bilayer graphene.** (A) A schematic of scanning photocurrent setup. The red region represents the hot carriers generated under the tip. (B) Top: Seebeck coefficient $S$ profile across a domain wall calculated from first principles (Supplementary Note 3 of (*15*)). The DW is located at $x = 0$. Middle: perspective view of the experiment showing photocurrent generation at the domain wall. The green background represents the Seebeck coefficient profile and the blue dots represent carriers generated by thermoelectric effect. Bottom: schematic of the BLG band structure across the DW for three different stackings AB, BA and saddle point (SP). (C) Photocurrent ($I_{PC}$) image taken with $\omega = 900\ \text{cm}^{-1}$ and $V_G = +24\ \text{V}$ at $T = 300\ \text{K}$. (D) Spatial gradient of the photocurrent defined as $dI_{PC}/dx$ of the data in (C). (E) Line profiles of $I_{PC}$ and $dI_{PC}/dx$ across a DW (shown as red and magenta lines in (C) and (D)).

In minimally twisted bilayer graphene (MTBG), the moiré pattern periodicity is large, e.g., 140 nm for $\theta \approx 0.01°$ and prone to atomic relaxation. In the relaxed state, the Bernal stacked domains (AB and BA) dominate while the less stable stacking configurations are reduced to a network of narrow domain walls (DWs). TEM measurements have shown that the DWs are 6-9 nm wide (*10*). Previous transport (*11*), nano-infrared (*12*, *13*), and STM (*14*) studies have revealed the existence of topological states at the DWs when an electronic bandgap is opened by a sufficiently large interlayer bias between the graphene sheets. At smaller interlayer biases, the change in the atomic stacking across the DW still leads to a change in the electronic properties.

Scanning nano-photocurrent imaging has emerged as a novel optoelectronic probe capable of resolving changes in DC transport properties of graphene with nanometer scale spatial resolution (*16*). Previous nano-photocurrent experiments have resolved charge inhomogeneities and grain boundaries in monolayer graphene (*16*) and mapped variations in twist angle of TBG at twist angles $\theta > 1^\circ$ (*17*). Here we use scanning nano-photocurrent imaging to study domain walls in MTBG. We show that the photocurrent patterns arise from DC Seebeck coefficient variations occurring at the DWs on a nanometer length scale. We further propose and demonstrate a mechanism that utilizes the intrinsic hyperbolicity of the hBN substrate to enhance the DW features in photocurrent images.

## Results and Discussion

Figure 1(A) shows a schematic of our experiment. Infrared light is focused onto the apex of a sharp metallic tip which enhances the electric field underneath the tip. The enhanced field locally generates a photocurrent which we collect through electrical contacts at zero bias. In graphene, the photocurrent arises from electronic inhomogeneities through the photothermoelectric effect (PTE), schematically shown in Figure 1(B) (*18–20*). Photocurrent images are acquired by raster scanning the tip across the sample. Our technique overcomes the diffraction limit and provides a spatial resolution of about 20 nm while also allowing for simultaneous nano-infrared imaging (*16*). Our device consists of two graphene layers with a minimal relative twist encapsulated between 37 nm bottom hexagonal boron nitride (hBN) layer and 6nm top hBN layer. The entire stack rests on a 285 nm $SiO_2$/Si substrate with the $SiO_2$ layer serving as the gate dielectric. Piezoresponse force microscopy (PFM) (*21*) before encapsulation of the device revealed domain walls with a periodicity of about 500 nm (Supplementary Note 1 of (*15*)).

Figure 1(C) shows a representative photocurrent image of our device acquired at room temperature with laser frequency of $\omega = 900\ \mathrm{cm}^{-1}$. We use a color scheme that enables easy identification of the sign of the photocurrent: red and blue represent positive and negative currents respectively while white represents regions where the measured current is zero, thus highlighting the zero-crossing contours. Some of the zero-crossing contours form easily identifiable lines in the $y$-direction while others form a meandering pattern. On closer inspection, we find a series of fine structures in the photocurrent image that form a hexagonal lattice. These features are more clearly revealed in the map of the photocurrent gradient, $dI_{PC}/dx$, shown in Figure 1(D). The periodicity of these features is consistent with the domain walls observed in PFM images before encapsulation (Supplementary Note 1 of (*15*)). In the $dI_{PC}/dx$ image, the vertical domain walls appear to be more intense because of the contact configuration used in our experiments, as explained in Supplementary Note 3.1 of (*15*). The lattice structure and the matching periodicity lead us to conclude that the fine features correspond to the domain walls of a relaxed moiré superlattice in TBG.

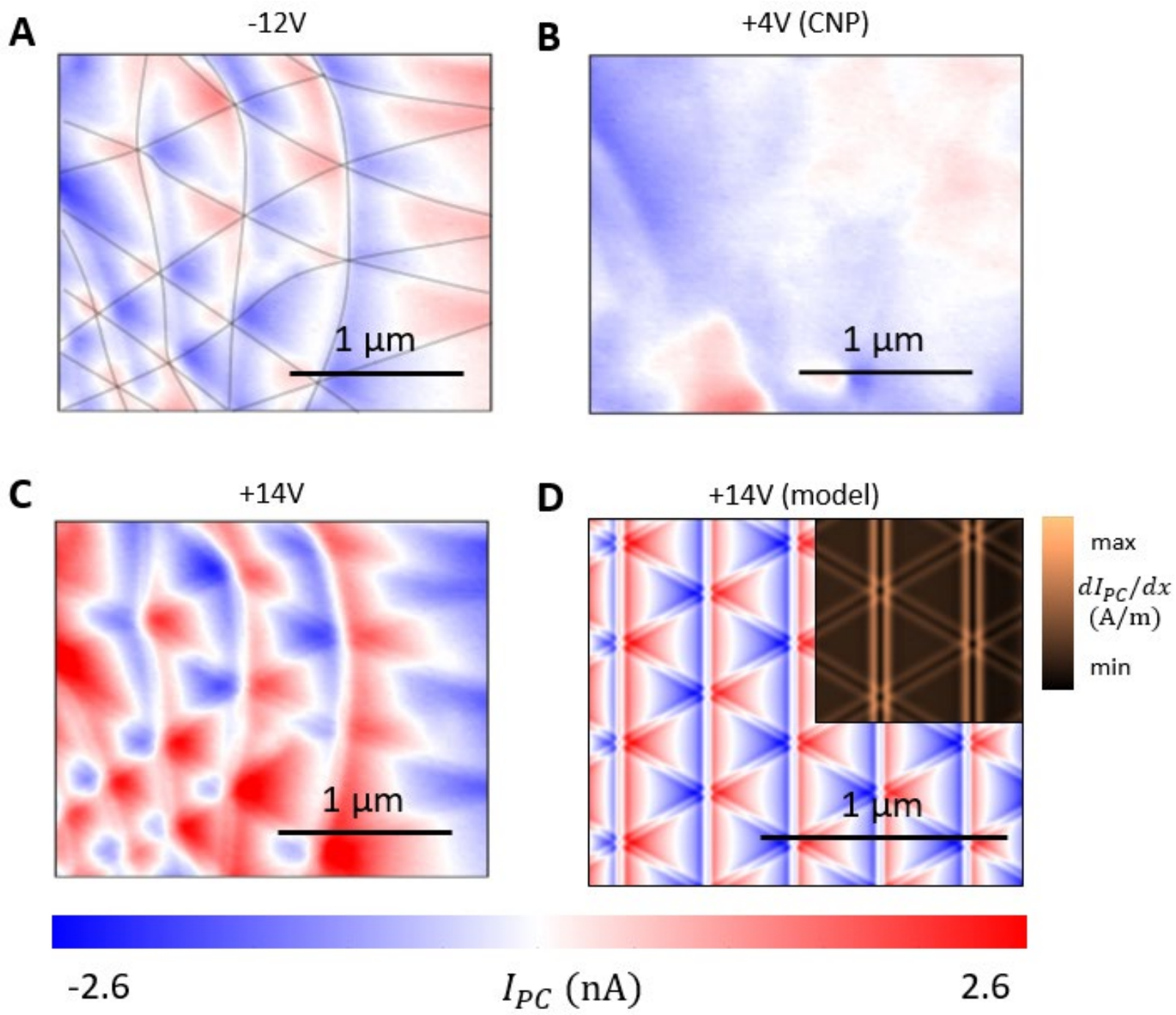

**Figure 2 | Thermoelectric origin of the photocurrent in TBG.** (A - C) Gate voltage dependence of the photocurrent ($I_{PC}$) at $\omega = 900\text{cm}^{-1}$ and $T = 300\text{K}$. Gate voltage is indicated above each panel. (D) Calculated photocurrent pattern using the Shockley-Ramo formalism (*22*) with material parameters corresponding to $V_G = +14\text{V}$ (more details in Supplementary Note 3 of (*15*)). The top-right inset shows the calculated image of $dI_{PC}/dx$ (compare with Figure 1(D)).

Next, we study the gate dependence of the photocurrent maps as plotted in Figure 2(A - C). Transport experiments on our device showed that the charge neutrality point (CNP), where the carrier density is minimal and the majority carriers change from holes to electrons, occurs at $V_G = +4$ V (Supplementary Note 1 of (*15*)). Photocurrent imaging at the CNP (Fig 2(B)) does not show any of the features observed in Fig 1(C). A comparison of the images at $V_G = -12$ V (Fig 2(A)) and $V_G = +14$ V (Fig 2(C)) reveals that the photocurrent has identical meandering pattern and fine DW features for positive and negative gate voltages except for a sign change. These results show that the meandering patterns and the DW features are antisymmetric with respect to the carrier type. As the gate voltage increases further in both the positive and negative direction, we find that the patterns weaken and eventually become unresolvable (Supplementary Note 2 of (*15*)). We note that the carrier densities in Figure 2 are too low to produce significant plasmonic effects in bilayer graphene (Supplementary Note 1.2 of (*15*)).

Previous theoretical (*20*) and experimental (*19*, *23*, *24*) investigations have discovered that the dominant mechanism for photocurrent generation in graphene is the photothermoelectric effect (PTE). In this mechanism, the absorption of incident light generates hot carriers in graphene. When the hot carriers encounter variations in the Seebeck coefficient, a thermoelectric voltage is generated which drives a current through the sample. The spatial profile of the measured current is therefore directly related to the Seebeck coefficient profile in the sample. PTE shows several characteristic features in experiments. First, since Seebeck coefficient is antisymmetric with respect to the sign of the carriers, the resulting photocurrent patterns also change sign when the carrier type changes from holes to electrons (*19*, *23*). Second, the Seebeck coefficient of bilayer graphene rapidly diminishes as the carrier density increases (*23*, *25*). Therefore, any variations in the Seebeck coefficient and the resulting photocurrent must also become small. Both features are present in our data, strongly suggesting that the photocurrent patterns we observe arise from PTE.

To confirm our hypothesis that the photocurrent arises from PTE and to gain a deeper understanding of our results, we calculated the expected photocurrent patterns from PTE. The input to these calculations are the Seebeck coefficient profile and the hot carrier temperature profile. We computed the former for an isolated domain wall using a generalized Boltzmann approach (Supplementary Note 3 of (*15*)) and the resulting profile is shown in Figure 1(B). To compare with our experiment, we superposed the one-dimensional Seebeck profiles in a hexagonal pattern to generate a two-dimensional lattice of domain walls (Supplementary Note 3.4 of (*15*)). Next, we computed the spatial profile of the hot carriers. We first computed the electric field at the graphene surface using two different models for the tip (1) a "lightning-rod model" in which the tip is represented by a conducting hyperboloid and (2) a simplified common approximation of the tip by a vertically oriented point dipole (Supplementary Note 3.3 and 3.4 of (*15*)). Since the conductivity of the graphene sheet is dominated by the in-plane components, we assumed that the radially symmetric in-plane field, $E_r$, governs the generation of hot carriers. We then solved the heat equation to determine the spatial profile of the hot carrier temperature (Supplementary Note 3.1 of (*15*)).

The Seebeck coefficient profile and the electron temperature profile are sufficient to calculate the local thermoelectric voltage for a given tip position. For gapless materials such as graphene, the photocurrent collected by distant electrodes also depends on the contact geometry. We used the Shockley-Ramo formalism of Ref (*22*) to include the effects of the contacts and our calculation procedures are described in more detail in Supplementary Note 3 of (*15*).

The photocurrent pattern resulting from the hyperboloid tip calculation is shown in Figure 2(D). Our results reproduce the key features of our data including the meandering patterns and the fine features at the domain walls. We can now correlate the features in the photocurrent images with those in the Seebeck coefficient. The fine features and the zero-crossing contours that form straight lines along the $y$-axis arise from the domain walls themselves. On the other hand, the meandering zero-crossing contours go across domain walls, and arise from the interference of photocurrents generated by neighboring domain walls. The good agreement between calculations and data confirms that our photocurrent experiments directly probe the nanometer-scale Seebeck coefficient variations present at the domain walls.

While the first-principles Seebeck coefficient profile produced a photocurrent pattern similar to the experiment, we note that our experiment is not sensitive to the fine details of the Seebeck coefficient at the domain wall. In fact, any change in Seebeck on a length scale significantly shorter than the spatial extent of the hot carriers (typically called the cooling length (*16*)) will produce a pattern similar to the experiment, as we demonstrate in Supplementary Note 3.3 of (*15*).

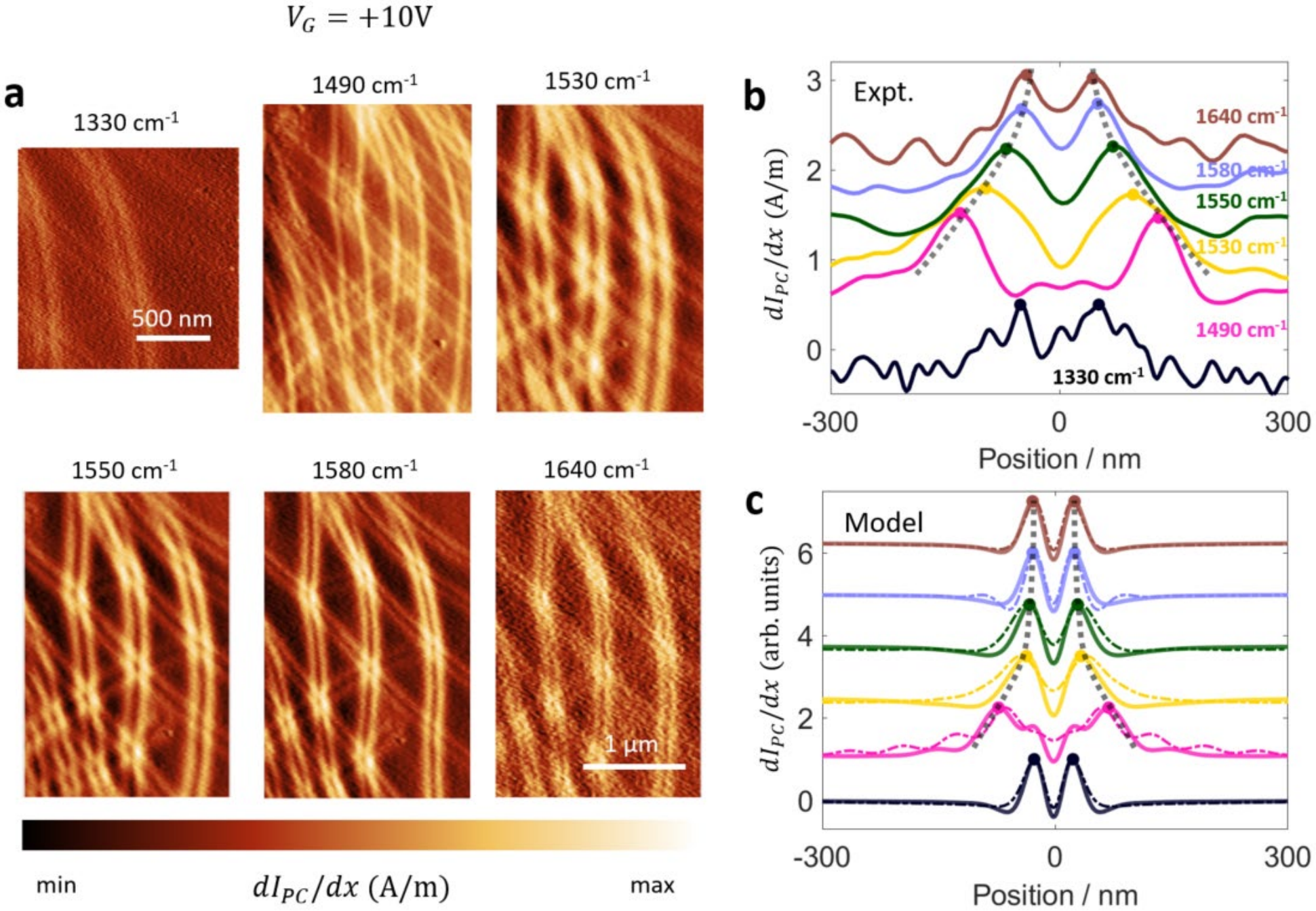

**Figure 3 | Domain wall photocurrent patterns in the hBN Reststrahlen band.** (A) Gradient of photocurrent ($dI_{PC}/dx$) for several frequencies around the hBN Reststrahlen band. (B) Experimental line profiles of $dI_{PC}/dx$ for several frequencies. The black dashed lines are guides to the eye. (C) Photocurrent profiles calculated using the frequency dependent electric field profiles. The thick solid lines correspond to the hyperboloid tip and the thin dashed lines correspond to the point dipole model. The theoretical curves are normalized to their respective maxima. Curves in panel (B) and (C) are offset vertically for clarity.

So far, the hBN layers which surround the graphene sheet have not played an active role. We now show that the optical properties of hBN can be exploited to enhance the photocurrent features from the DWs. Over two frequency bands in the mid-infrared, referred to as the lower and upper Reststrahlen bands, the permittivity of hBN along its in-plane and out-of-plane principal axes have opposite signs (*26*). Such behavior, known as hyperbolicity, leads to highly confined phonon polaritons (*26*–*30*) and hyperlensing effects (*31*, *32*). Here, we specifically focus on the upper Reststrahlen band (1376 to 1614 cm$^{-1}$) where hBN transverse dielectric constant in the $xy$-plane becomes negative ($\epsilon_t < 0$). The out-of-plane dielectric constant remains positive ($\epsilon_z > 0$) and is weakly frequency dependent.

We performed photocurrent experiments at several frequencies around the upper Reststrahlen band and the data is shown in Figure 3(A). We observe a clear change in the width of the domain wall feature with frequency. Specifically, we find that at the lower end of the Reststrahlen band (e.g., $\omega = 1490\ \mathrm{cm}^{-1}$ and $\omega = 1530\ \mathrm{cm}^{-1}$ in Fig. 3(A)) the domain wall feature is wider compared with pattern below the reststrahlen band (compare, for example, with $\omega = 900\ \mathrm{cm}^{-1}$ of Figure 1(D)). As the frequency increases, the width of the broad features decreases. Finally, at frequencies above the Reststrahlen band ($\omega = 1640\ \mathrm{cm}^{-1}$ in Fig. 3(A)), the width of the feature returns to its value below the Reststrahlen band. Furthermore, at $\omega = 1490\mathrm{cm}^{-1}$ we observe two faint peaks in between the two stronger peaks. These effects are further confirmed by the frequency-dependent line profiles shown in Fig 3(B). From the line profiles, we see that the fainter peaks at $\omega = 1490\mathrm{cm}^{-1}$ are approximately coincident with the original peaks at $\omega = 1330\mathrm{cm}^{-1}$ and $1640\mathrm{cm}^{-1}$.

Since our experiments at $\omega = 900\ \mathrm{cm}^{-1}$ and the related modelling have shown that the photocurrent pattern is of PTE origin, any change in the pattern must be due to either a change in the Seebeck coefficient profile or the hot carrier profile. The Seebeck coefficient is not expected to change with the frequency of light incident on the material in the linear regime and the laser power used in our experiment (~20mW, see Supplementary Note 1 of (*15*)) is too weak to produce a significant non-linear effect. Therefore, we are led to conclude that change in the hot carrier distribution must be responsible for the observed change in width.

The spatial profile of Joule heating power is determined by the electric field profile under the tip and the real part of the optical conductivity of bilayer graphene, $\mathrm{Re}(\sigma)$. The frequency dependence data of Figure 3 was collected at $V_G = +10\mathrm{V}$, where the estimated Fermi energy in the Bernal stacked regions is low ($E_F \approx 10\mathrm{meV}$, refer to Supplementary Note 1.2 of (*15*)) and the optical conductivity is dominated by the frequency-independent interband conductivity (*33*, *34*). Therefore, we conclude that the electric field profile under the tip must change with frequency within the Reststrahlen band in order to reproduce the experimental observations shown in Figure 3. To model the observed change in width, we used the “lightning rod” model and a point dipole model to compute the radial electric field at several frequencies around the Reststrahlen band (Supplementary Note 3.3 and 3.4 of (*15*)). The photocurrent profiles from our modeling are shown in Figure 3(C) and show good agreement with the experiment.

The electric field at the graphene layer can be thought of as the sum of two separate parts. The first part is the incident field from the tip and the second part is the field reflected by the hBN substrate in response to the tip excitation. The left panels in Figure 4(A) show the tip field and the right panels show the total field. We see that the tip field is weakly dependent on the frequency but the field reflected by the substrate is strongly modified inside the Reststrahlen band. The wider electric field leads to a wider hot carrier temperature profile (Figure 4(B)) and a broader photocurrent pattern (Figure 3(C)).

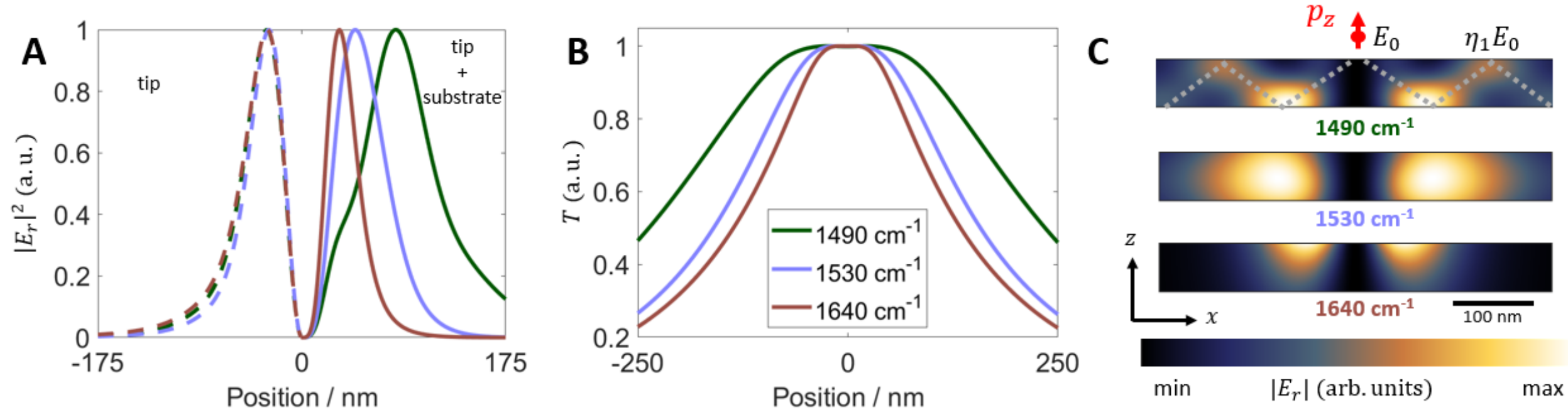

**Figure 4 | Local electric field and temperature inside and outside the Reststrahlen band.** (A) In-plane electric field $|E_r|^2$ at the graphene layer calculated for a hyperboloid tip. The left half (dashed lines) shows the field of the tip alone and the right half (solid lines) shows the total field from the tip and the substrate. (B) Hot carrier temperature $T$ calculated using the total field from (A). (C) Cross section of a hBN slab showing the electric field resulting from excitation by a point dipole located above the hBN surface. 1490 cm$^{-1}$ and 1530 cm$^{-1}$ are inside and 1640 cm$^{-1}$ is outside the Reststrahlen band, respectively. The dashed line in the 1490 cm$^{-1}$ image shows the polariton propagation. $E_0$ and $\eta_1 E_0$ represent the magnitude of the in-plane field at the zeroth order and the first order maxima respectively.

The origin of this widening is closely related to a previously observed effect in hBN slabs, known as hyperlensing (*31*, *32*, *35*). In hyperlensing, a sub-wavelength antenna launches phonon polariton rays that propagate inside the hBN slab. Here, our tip acts as the antenna. The total in-plane field at the hBN surface can be thought of as a series of concentric rings centered below the tip with a radius of $r_k^{max}$ for the $k$-th ring. The electric field at the $k$-th ring is given by $E_k = \eta_k E_0$, where $\eta_k$ is related to the permittivity of the hBN slab and the substrate (Supplementary Note 3.4 of (*15*)). The $k = 0$ ring corresponds to the field from the tip itself with magnitude $E_0$ while $k = 1, 2, \ldots$ correspond to phonon polaritons propagating in the hBN slab (see Fig 4(C) and Supplementary Note 3.4 of (*15*)). Therefore, inside the Reststrahlen band, the zeroth order maximum is frequency independent but the higher order maxima are strongly frequency dependent.

Typically, the magnitude of the field at the $k = 1$ ring is expected to be smaller than the field created directly by the tip ($|\eta_1| < 1$). However, for several frequencies inside the Reststrahlen band, $|\eta_1| > 1$, so $E_1$ dominates and leads to a broad frequency-dependent electric field profile and photocurrent pattern. The faint central features in $dI_{PC}/dx$ at $\omega = 1490\text{cm}^{-1}$ can now be understood as arising from $E_0$ while the stronger broader features arise from $E_1$. In principle, the polariton reflections corresponding to $E_2$ and higher order terms should be reflected in the photocurrent profile. Our simulations suggest that a sharper tip and more widely separated domain walls (i.e., smaller twist angle) could reveal such features in future photocurrent experiments (Supplementary Note 3.4 of (*15*)).

In conclusion, we have demonstrated that nano-photocurrent experiments are sensitive to nanoscale changes in the Seebeck coefficient at the domain walls in MTBG. Our modeling of the photocurrent patterns is consistent with experiment. We further demonstrate a hyperbolic optoelectronic effect where the domain wall photocurrent patterns are enhanced by the hyperbolicity of the hBN substrate.

Note: While preparing our manuscript for submission, we became aware of a similar work by Hesp et al (*36*).

## Methods

### *Device fabrication*

The minimally twisted bilayer graphene device was fabricated using the dry transfer method. Piezoresponse force microscopy (PFM) (*21*) was performed before encapsulation to ensure that a moiré pattern with a large periodicity was present (Supplementary Figure 1(A) of (*15*)). The contact geometry was specifically designed for easy interpretation of photocurrent experiments (Supplementary Figure 1(B) of (*15*), refer to Supplementary Note 3.1 of (*15*) on photocurrent modeling). We used the M1-M3 contacts for all photocurrent experiments.

### *Bilayer graphene parameter estimate*

The properties of bilayer graphene depend not only on the carrier density but also on the interlayer bias. In our experiment, we have a single Si back gate which allows us to control the carrier density accurately. Here, we describe our estimate of the interlayer bias values for different gate voltages.

First, we assume that the interlayer bias is zero at charge neutrality point $V_G = +4\text{V}$. This assumption is reasonable for the ultra-high quality, doubly-encapsulated devices studied in this work (*37*). For a given gate voltage, we can directly calculate the displacement field below the graphene layers:

$$D_{lower} = \frac{\epsilon_{lower} V_G}{d_{lower}} \qquad \text{(Eq 1)}$$

where $\epsilon_{lower}$ and $d_{lower}$ are the dielectric constant and thickness of the $SiO_2$ dielectric layer. Because we have no top gate, the displacement field above the graphene layers $D_{upper} = 0$ and effective displacement field across the graphene is given by:

$$\overline{D} = \frac{D_{upper} + D_{lower}}{2} = \frac{D_{lower}}{2}. \qquad \text{(Eq 2)}$$

We use Ref (*38*) to estimate the interlayer bias $V_i$ from $\overline{D}$. To estimate $E_F$, we keep $V_i$ fixed and vary the Fermi energy $E_F$ until the carrier density we calculate with a tight-binding model matches the value expected from capacitance calculations. Supplementary Figure 2 of (*15*) shows a plot of the estimated $E_F$ and $V_i$ for several gate voltages. We find that the estimated Fermi energy is linear with gate voltage. At small displacement fields, the band structure of bilayer graphene can be well approximated to be parabolic (*39*). In 2 dimensions, a parabolic dispersion leads to a constant density of states and a linear dependence of the Fermi energy on carrier density, which is consistent with our estimate. We note that the carrier densities considered in this manuscript (Figures 2 and 3) are too low to produce significant plasmonic effects. In bilayer graphene, plasmons are typically observed in nano-infrared imaging for $V_G - V_{CNP} > \sim 30\text{V}$ (*40*).

### *Nano-photocurrent experiments*

Room temperature nano-photocurrent measurements were performed in a commercial s-SNOM from Neaspec GmbH. Low temperature nano-photocurrent measurements were performed in a home-built SNOM within an ultrahigh vacuum chamber (*41*) at $T = 200\text{K}$. For the $\omega = 900\text{cm}^{-1}$ experiments, we used a $CO_2$ laser and for the Reststrahlen band experiments, we used a tunable quantum cascade laser from Daylight Solutions. The incident laser power was around 20mW in all cases. The current was measured using a Femto DHPCA-100 current amplifier. To isolate the photocurrent contributions from the near-fields localized under the tip, the measured current was demodulated at a harmonic $n$ of the tapping frequency. In this work, we used $n = 3$ for room temperature experiments and $n = 2$ for low temperature experiments.

## Acknowledgements

Research in van der Waals heterostructures at Columbia was solely supported as part of Programmable Quantum Materials, an Energy Frontier Research Center funded by the U.S. Department of Energy (DOE), Office of Science, Basic Energy Sciences (BES), under award DE-SC0019443. D.N.B. is the Vannevar Bush Faculty Fellow (N00014-19-1-2630) and Moore Investigator in Quantum Materials EPIQS #9455. A.R. and M.M.F. were supported by The Office of Naval Research under grant N00014-18-1-2722. D.H. is supported by a grant from the Simons Foundation (579913). T.S. is supported by Spain's MINECO under Grant No. FIS2017-82260-P as well as by the CSIC Research Platform on Quantum Technologies PTI-001.

## Author contributions

SC and DEG-A fabricated the MTBG device under the supervision of JCH and CRD. DH and SSS performed the nano-photocurrent experiments and analyzed the data with assistance from ASM, AR and MM. TS performed the Seebeck coefficient calculations. ASM, AR, MEB, CFBL and MM provided the electromagnetic simulations. SSS and DH wrote the manuscript with inputs from all authors. DNB supervised the entire effort.

## Competing Interests

The authors declare no competing interests

# Supplementary Information for "Hyperbolic enhancement of photocurrent patterns in minimally twisted bilayer graphene"

S. S. Sunku[1,2,*], D. Halbertal[1,*,†], T. Stauber[3], S. Chen[1], A. S. McLeod[1], A. Rikhter[4], M. E. Berkowitz[1], C. F. B. Lo[1], D. E. Gonzalez-Acevedo[1], J. C. Hone[5], C. R. Dean[1], M. M. Fogler[4], D. N. Basov[1]

[1] Department of Physics, Columbia University, New York, NY

[2] Department of Applied Physics and Applied Mathematics, Columbia University, New York, NY

[3] ICMM at CSIC, Madrid, Spain

[4] Department of Physics, University of California, San Diego, La Jolla, CA

[5] Department of Mechanical Engineering, Columbia University, New York, NY

* These authors contributed equally

† dh2917@columbia.edu (D.H.)

## Supplementary Note 1: Device characterization

Supplementary Figure 1 shows the piezoresponse force microscopy (PFM) (*1*) image of the graphene layers before encapsulation, the contact configuration used for photocurrent experiments and the determination of charge neutrality. Supplementary Figure 2 shows the dependence of the bilayer graphene parameters $E_F$ and $V_i$ on the applied gate voltage.

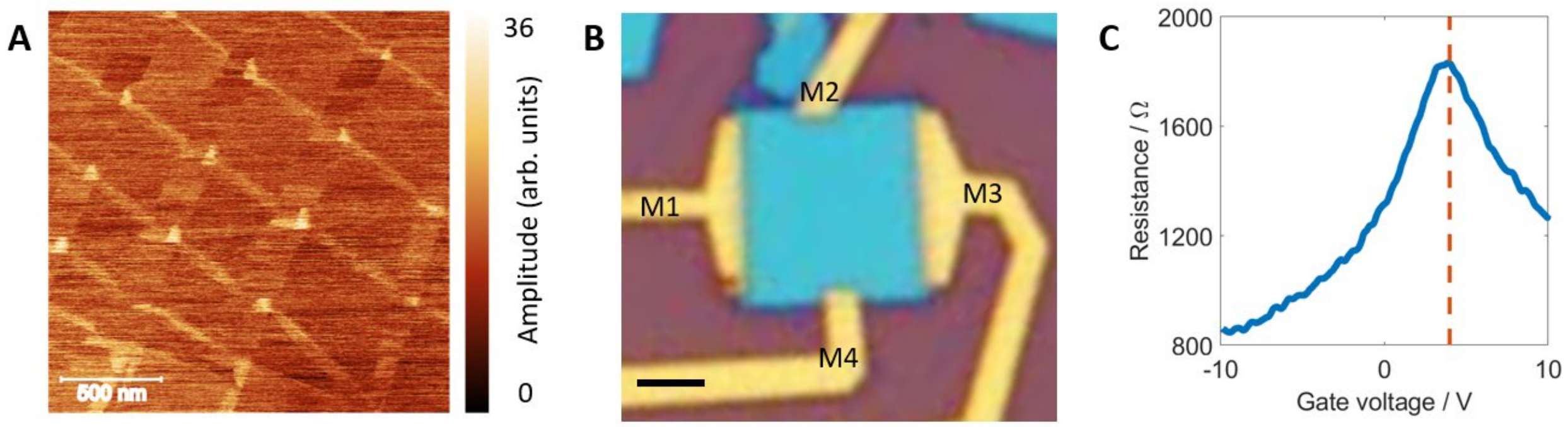

**Supplementary Figure 1 | Device fabrication.** (A) Piezoresponse force microscopy image of the graphene layers before encapsulation showing domain walls. (B) Optical microscope image showing the final contact configuration. Scale bar $3\mu\mathrm{m}$. (C) Two probe resistance measured using M1 and M3 contacts as a function of $V_G$ applied to the Si back gate. The dashed line corresponds to $V_G = +4\mathrm{V}$ which is taken to be the charge neutrality point (Figure 2 (A) of main text).

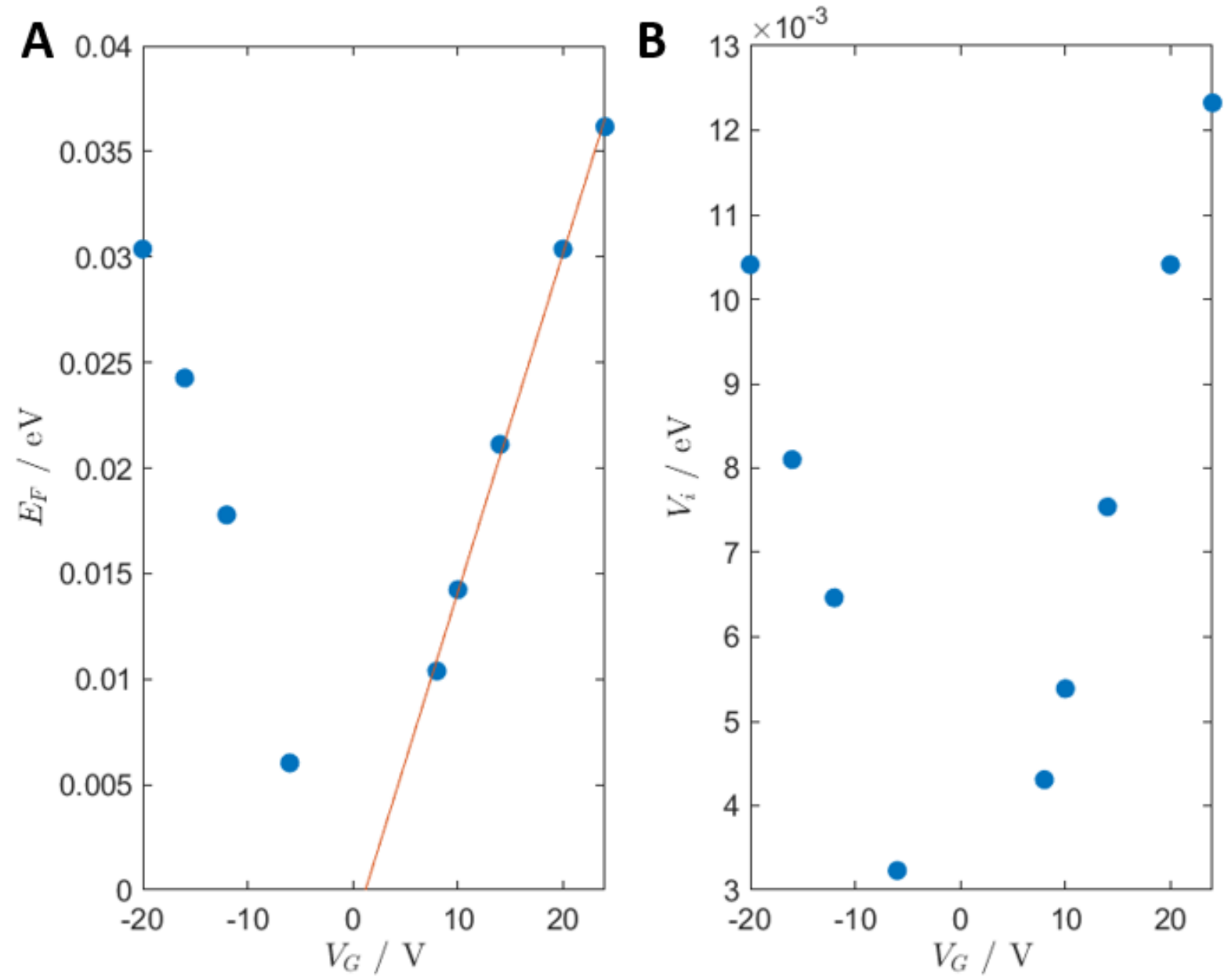

**Supplementary Figure 2 | Estimated Fermi energy and interlayer bias for bilayer graphene with a single gate.** The dots correspond to specific gate voltages and the red line is a linear fit.

## Supplementary Note 2: More photocurrent data

Here, we describe our analysis methods for the photocurrent data and include all of the collected images. The photocurrent signal was demodulated at a harmonic of the tip tapping frequency with a lock-in amplifier. The phase offset of the demodulation signal is arbitrary since the phase only determines the direction of the current and otherwise does not contain any meaningful information. Therefore, for each photocurrent image, we adjusted the phase offset so as to maximize the signal in the in-phase component and minimize it in the out-of-phase component. Stated more rigorously, $S_{in}(x,y), S_{out}(x,y)$ are the raw data images for in-phase and out-of-phase lock-in output channels. For an offset phase $\phi_0$, the corrected signal $S'_{in}(x,y), S'_{out}(x,y)$ is the result of rotation by $\phi_0$:

$$\begin{pmatrix} S'_{in} \\ S'_{out} \end{pmatrix} = \begin{pmatrix} \cos\phi_0 & \sin\phi_0 \\ -\sin\phi_0 & \cos\phi_0 \end{pmatrix} \begin{pmatrix} S_{in} \\ S_{out} \end{pmatrix} \qquad \text{(Eq 1)}$$

The offset angle $\phi_0$ is chosen as to minimize the variance of $S'_{out}$ across the image.

### Supplementary Note 2.1: $\omega = 900\text{cm}^{-1}$

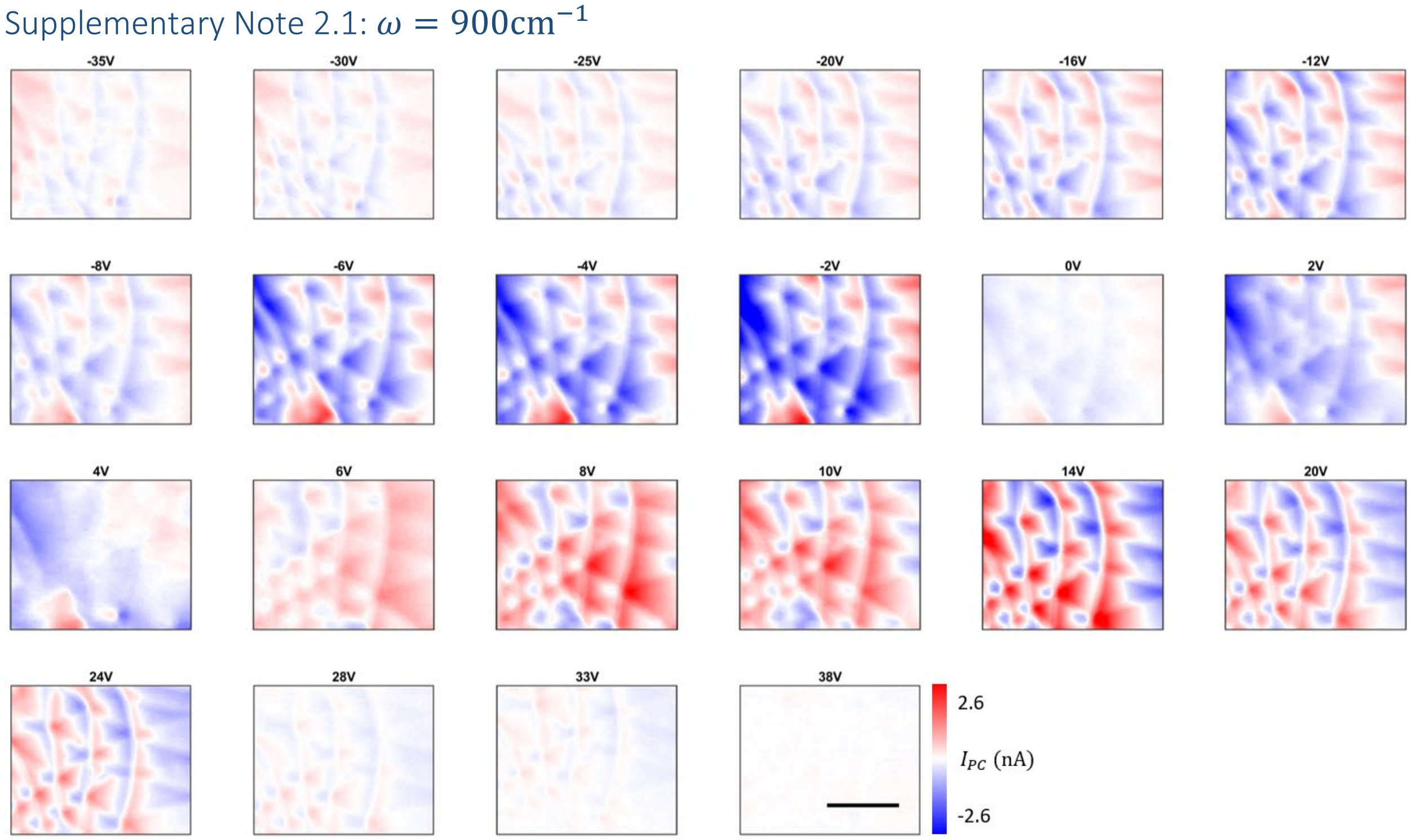

**Supplementary Figure 3 | Photocurrent data for several gate voltages at $\omega = 900\text{cm}^{-1}$.** Scale bar 1μm.

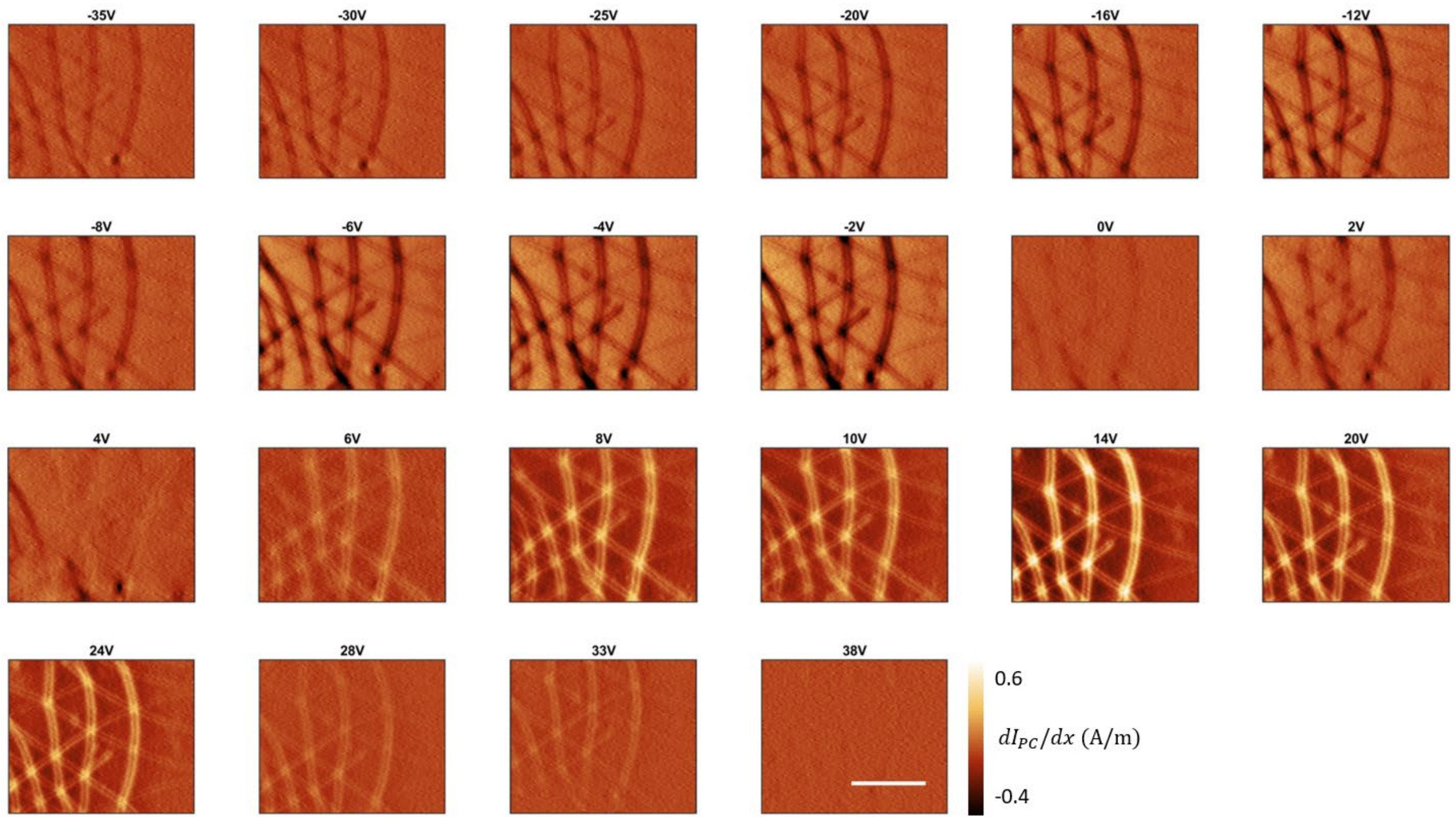

**Supplementary Figure 4 | Photocurrent gradient for several gate voltages at $\omega = 900\mathrm{cm}^{-1}$.** Scale bar 1μm.

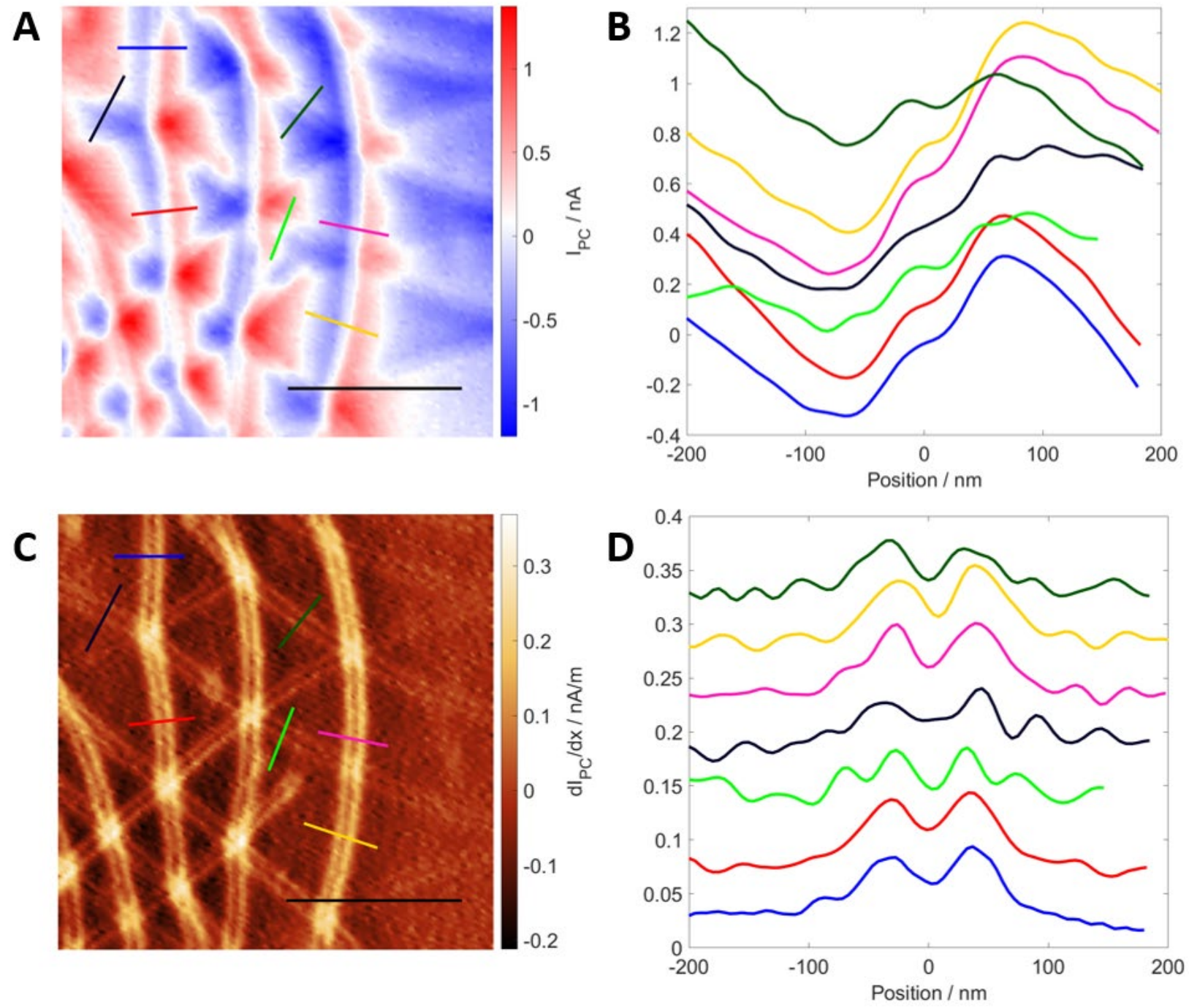

**Supplementary Figure 5 | Photocurrent line profiles at $V_G = +24\mathrm{V}$.** (A) Nano-photocurrent image at $V_G = +24\mathrm{V}$ (same as Figure 1(B) of main text). (B) Multiple line profiles across the domain walls. Each profile is offset by an arbitrary number for clarity. (C) and (D) same as (A) and (B) but for $dI_{PC}/dx$.

## Supplementary Note 2.2: hBN reststrahlen band

**Supplementary Figure 6 | Full frequency dependent plots of the photocurrent in the hBN reststrahlen band at $V_G = +10\text{V}$.**

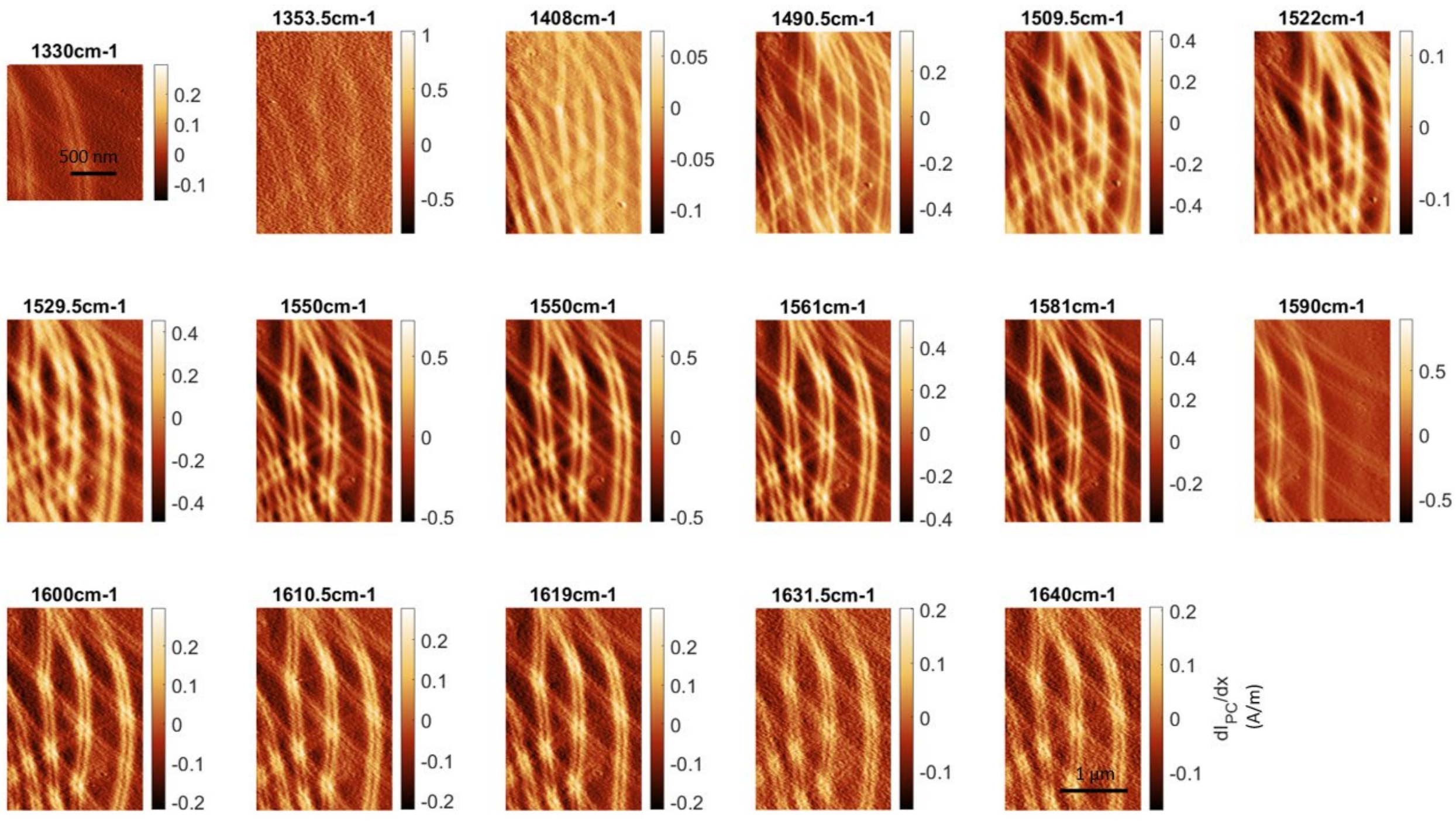

**Supplementary Figure 7 | Frequency dependent plots of the gradient of the photocurrent in the hBN reststrahlen band at $V_G = +10\text{V}$.**

## Supplementary Note 3: Photocurrent model

### Supplementary Note 3.1: Photocurrent calculation

In gapless materials such as graphene, the spatial photocurrent profiles are described by the Shockley-Ramo formalism (*2*). In this formalism, an auxiliary potential $\phi$ is defined as solution of Laplace's equation, $\nabla \cdot (\sigma^T \nabla \phi) = 0$ ($\sigma$ is the dc conductivity tensor) with the contact configuration dependent boundary conditions: $\phi = 1$ at current collecting contacts (where the current is being measured) and $\phi = 0$ at the rest of the grounded contacts. According to the Shockley-Ramo formalism, one can show that the measured photocurrent would then be:

$$I_{PC} = \iint d^2\boldsymbol{r}' \boldsymbol{J}_{local}(\boldsymbol{r}') \cdot \nabla\phi(\boldsymbol{r}') \qquad \text{(Eq 2)}$$

Where $\boldsymbol{J}_{local}$ is the locally generated photocurrent density. In our case the photocurrent is generated through the photothermoelectric effect, and for a tip positioned at a point $\boldsymbol{r}$ would therefore yield the following photocurrent reading:

$$I_{PC}(\boldsymbol{r}) = \iint d^2\boldsymbol{r}'\, \sigma(\boldsymbol{r}')\, S(\boldsymbol{r}')\, \nabla T(\boldsymbol{r}', \boldsymbol{r}) \cdot \nabla\phi(\boldsymbol{r}') \qquad \text{(Eq 3)}$$

where $T(\boldsymbol{r}', \boldsymbol{r})$ is the temperature at $\boldsymbol{r}'$ as a result of a tip located at $\boldsymbol{r}$ and $S$ is the Seebeck coefficient tensor.

Since we are interested in a 1D domain wall, we can simplify the problem with a quasi-1D geometry. We assume that the sample is infinite in the $y$ direction, both $\sigma$ and $S$ are independent of $y$, and we have a grounded contact at $x = 0$ and a collecting contact at $x = L$. We further assume that $S$ is diagonal and isotropic. These assumptions yield: $\phi(x) = \int_0^x dx' \frac{1}{\sigma(x')} / \int_0^L dx' \frac{1}{\sigma(x')}$. After substitution into the photocurrent expression we get:

$$I_{PC}(\boldsymbol{r}) = \frac{\Sigma}{L} \iint d^2\boldsymbol{r}'\, S(\boldsymbol{r}') \frac{\partial T(\boldsymbol{r}', \boldsymbol{r})}{\partial x} \qquad \text{(Eq 4)}$$

Where $\Sigma \equiv L \left( \int_0^L dx' \frac{1}{\sigma_{xx}(x')} \right)^{-1}$. Finally, we assume the shape of the temperature profile to be independent of tip position, such that: $T(\boldsymbol{r}', \boldsymbol{r}) = T(\boldsymbol{r}' - \boldsymbol{r})$. This assumption is justified if the absorption and thermal properties are not strongly modulated as a function of position. The last assumption formulates the above expression for the measured photocurrent as a 2D convolution of two terms such that:

$$I_{PC}(\boldsymbol{r}) = \frac{\Sigma}{L} \left( S * \frac{\partial T}{\partial x} \right)(\boldsymbol{r}) \qquad \text{(Eq 5)}$$

The remaining task in order to calculate the photocurrent is to calculate the temperature spatial profile, $T(\boldsymbol{r})$. We describe it by the diffusion equation:

$$-\kappa\nabla^2\tau(\boldsymbol{r}) + g\tau(\boldsymbol{r}) = P(\boldsymbol{r}) \quad \text{(Eq 6)}$$

where $\tau = T - T_0$ is the electronic temperature change relative to a background thermal bath at $T_0$, $\kappa$ is the in-plane thermal conductivity of graphene, $g$ is the out-of-plane thermal coupling to the substrate (both assumed to be spatially uniform for simplicity) and $P$ is the absorbed heat distribution (which is estimate in this study using the lightning rod model as described in Supplementary Note 3.3). Following (*3*), the general solution can be obtained by a Green's function approach, where we first solve for the Green's function $G$ that satisfies the impulse response equation:

$$-\kappa\nabla^2 G(\boldsymbol{r}) + gG(\boldsymbol{r}) = \delta^{(2)}(\boldsymbol{r}) \quad \text{(Eq 7)}$$

where $\delta^{(2)}(\boldsymbol{r})$ is the 2D delta function. The general solution to Eq 6 for an arbitrary $P(\boldsymbol{r})$ is then given by the convolution $\tau = G * P$. We can solve for the Green's function through a Fourier analysis. We define $\tilde{G}(k_x, k_y) = \int_{-\infty}^{\infty} dx \int_{-\infty}^{\infty} dy\, G(x,y)e^{-i(k_x x + k_y y)}$ to be the Fourier transform of $G(x,y)$. One can then show that:

$$\tilde{G}(k_x, k_y) = \frac{1}{4\pi^2}\frac{1}{g + \kappa(k_x^2 + k_y^2)} \quad \text{(Eq 8)}$$

Taking the inverse Fourier transform gives us the Green's function

$$G(\boldsymbol{r}) = K_0\left(\frac{r}{\sqrt{\kappa/g}}\right) \quad \text{(Eq 9)}$$

where $K_0(x)$ is the 0$^{th}$ order modified Bessel function of the second kind and $l_{cool} = \sqrt{\kappa/g}$ is a thermal length-scale which is typically called the cooling length. In our simulations, we used $l_{cool} = 100\text{nm}$ for the room temperature $\omega = 900\text{cm}^{-1}$ data (Fig 2D) and $l_{cool} = 200\text{nm}$ for the $T = 200\text{K}$ data in the hBN Reststrahlen band (Figure 3C and 4B).

Assuming the graphene sheet thermal conductivity of $\kappa{\sim}10^{-6}\ \text{W/K}$ (*4*) and a room temperature cooling length of $l_{cool} = 100\ \text{nm}$, the interfacial thermal resistance in our samples is about $R_T = 1/g = l_{cool}^2/\kappa \sim 10^{-8}\ \text{m}^2\text{K/W}$, comparable to theoretically predicted (*5*) and experimentally measured (*6*) values. Note that this parameter appears to strongly depend on the interface and sample quality. In a previous photocurrent experiment (*7*), this thermal resistance was estimated to be as high as $10^{-5}\ \text{m}^2\text{K/W}$.

The Shockley-Ramo formalism also provides an explanation for the asymmetry between the domain wall $dI_{PC}/dx$ profiles along different directions in Figure 1. The profiles along the $y$-direction are significantly stronger because of the $\partial T/\partial x$ term in Eq 5. The domain walls along the other directions contribute less to the convolution in Eq 5 and therefore appear weaker in the experiment. This behavior is captured directly in Fig 2(D).

## Supplementary Note 3.2: First principles calculations of Seebeck coefficient across the domain wall

We will analyze the static transport properties across a single AB/BA domain wall. The Hamiltonian is adopted from (*8*) where the optical properties across a single domain wall were discussed i.e., we consider the general Hamiltonian of bilayer graphene

$$H = \begin{pmatrix} H_0 & U^\dagger \\ U & H_0 \end{pmatrix}, U = \begin{pmatrix} U_{AA} & U_{AB} \\ U_{BA} & U_{BB} \end{pmatrix}, \qquad \text{(Eq 10)}$$

where $H_0 = \hbar v_F \sigma \cdot \boldsymbol{k}$ denotes the Hamiltonian of a single layer graphene and $U$ the interlayer coupling with $U_{AA} = U_{BB} = \frac{t_1}{3}\left[1 + 2\cos\left(\frac{2\pi}{3}\frac{\delta}{a_0}\right)\right], U_{AB} = \frac{t_1}{3}\left[1 + 2\cos\left(\frac{2\pi}{3}\left(\frac{\delta}{a_0} + 1\right)\right)\right], U_{BA} = \frac{t_1}{3}\left[1 + 2\cos\left(\frac{2\pi}{3}\left(\frac{\delta}{a_0} - 1\right)\right)\right]$ (*9*). A single AB-BA domain wall at $x = 0$ with width $w$ is then modeled by the displacement field $\delta(x) = \frac{2}{\pi}\arctan\left[\exp\left(\frac{\pi x}{w}\right)\right] + 1$. For numerical convenience, we add another, independent, single BA/AB domain wall in order to implement periodic boundary conditions.

The particle current and heat-flow due to electrons is given by (*10*)

$$\begin{pmatrix} \vec{J} \\ \vec{U} \end{pmatrix} = \begin{pmatrix} \mathbf{K_0} & \mathbf{K_1} \\ \mathbf{K_1} & \mathbf{K_2} \end{pmatrix} \begin{pmatrix} e\vec{\nabla}\phi \\ T^{-1}\vec{\nabla}T \end{pmatrix} \qquad \text{(Eq 11)}$$

where the tensors $\mathbf{K}_l$ with $l = 0, 1, 2$ read

$$\mathbf{K}_l = \frac{g_s g_v}{A}\sum_{\boldsymbol{k},n} \vec{v}_{\boldsymbol{k},n}\, \vec{v}_{\boldsymbol{k},n}^T \tau_{\boldsymbol{k},n}\left(\epsilon_{\boldsymbol{k},n} - \mu\right)^l \left(-\frac{\partial f_{\boldsymbol{k},n}^0}{\partial \epsilon_{\boldsymbol{k},n}}\right). \qquad \text{(Eq 12)}$$

These quantities depend on the relaxation time $\tau_{\boldsymbol{k},n}$ and $\vec{v}_{\boldsymbol{k},n} = \langle \boldsymbol{k}, n | \hat{\vec{v}} | \boldsymbol{k}, n \rangle$ where $\epsilon_{\boldsymbol{k},n}$ and $|\boldsymbol{k}, n\rangle$ denote the eigenvalues and eigenvectors of the underlying Hamiltonian, respectively with $\boldsymbol{k}$ inside the first Brillouin zone. Furthermore, $f_{\boldsymbol{k},n}^0$ denotes the Fermi-Dirac distribution function at chemical potential $\mu$, $A$ denotes the area of the sample, $g_s = g_v = 2$ the spin and valley degeneracy, and $\hat{\vec{v}}$ is the velocity operator. Typical transport properties such as the dc conductivity, the Seebeck coefficient and the thermal conductivity are then defined by $\sigma_{dc} = e^2\mathbf{K}_0$, $S = -(eT)^{-1}\mathbf{K}_0^{-1}\mathbf{K}_1$, and $\kappa = T^{-1}(\mathbf{K}_2 - \mathbf{K}_1\mathbf{K}_0^{-1}\mathbf{K}_1)$.

Eq 11 can be generalized to define the local current response i.e., $\mathcal{J}(\vec{r}) = \int d\vec{r}'\mathcal{K}(\vec{r}, \vec{r}')\nabla\chi(\vec{r}')$ with $\mathcal{J}(\vec{r}) = \left(\vec{J}(\vec{r}), \vec{U}(\vec{r})\right)^T$ and the corresponding definitions for $\mathcal{K}(\vec{r}, \vec{r}')$ and $\chi(\vec{r})$. We then applied the local approximation (*11*) which amounts to $\mathcal{K}_{loc}(\vec{r}) = \int d\vec{r}'\mathcal{K}(\vec{r}, \vec{r}')$ and obtained the local transport quantities such as the Seebeck coefficient that were discussed in the main text.

## Supplementary Note 3.3: Electric field profiles using the lightning rod model

The electric field relevant for calculating a temperature profile at the graphene layer is computed using the lightning rod of probe-sample near-field interaction (*12*). Here the near-field probe is considered as an ideally conducting metallic hyperboloid (roughly conical in shape) 19 microns in height with a taper angle of about 20 degrees to the probe axis, and a curvature radius of 75 nm at its apex. For a chosen sample configuration comprising a multi-layer stack (here a 7 nm top hBN layer, nearly charge-neutral graphene bilayer atop a 36 nm hBN slab over an $SiO_2$ substrate), a specified probe-sample distance *d,* and illumination energy, the model predicts the axisymmetric charge distribution $\lambda(z) \equiv dQ/dz$ along the probe. For the ideally conducting probe, this charge conforms to the external profile of the probe in a quasi-continuum of rings of radius $\mathcal{R}(z)$, where $z$ denotes the probe's axial coordinate. From $\lambda(z)$, we evaluate the electric near-field from the probe in the graphene plane using the angular spectrum representation:

$$\boldsymbol{E_{probe}}(\rho, d) = \int_0^L dz\, \lambda(z) \int dq\, q\, [J_0(q\rho)\hat{z} + J_1(q\rho)\hat{\rho}]\, J_0(q\mathcal{R}(z)) e^{-q(d+z)} \qquad \text{(Eq 13)}$$

Here $\rho$ denotes the in-plane radial coordinate from the probe axis, $q$ is a Fourier momentum. As an integral sum of Bessel functions $J_1(q\rho)$, the radial field $E_{\rho,probe}$ presents a roughly "donut"-shaped in-plane distribution as shown in Fig. 4 of the main text. The total field inclusive of fields reflected from the sample is then given similarly by:

$$\boldsymbol{E_{total}}(\rho, d) = \int_0^L dz\, \lambda(z) \int dq\, q \begin{bmatrix} \left(1 + r_p(q)\right) J_0(q\rho)\hat{z} + \\ \left(1 - r_p(q)\right) J_1(q\rho)\hat{\rho} \end{bmatrix} J_0(q\mathcal{R}(z)) e^{-q(d+z)} \qquad \text{(Eq 14)}$$

Here $r_p(q)$ denotes the momentum-resolved Fresnel reflection coefficient for *p*-polarized fields computed for our heterostructure with a transfer matrix method.

We now turn our attention to the electric fields associated with generating the temperature profile relevant for the PTE underlying our photocurrent imaging. Since photocurrents were obtained at the $n = 2, 3$ harmonics of the probe tapping frequency $\Omega$, the spatially-resolved distribution of thermal power deposited in the graphene at these harmonics is given by:

$$P_n(\rho) \approx \mathrm{Re}(\sigma) \left|E_{\rho,n}\right|^2 \qquad \text{(Eq 15)}$$

Here $\sigma$ represents the optical conductivity of graphene and $E_{\rho,n}$ denotes the radially polarized total field demodulated at harmonic $n$:

$$E_{\rho,n}(\rho) \equiv \frac{\Omega}{\pi} \int_0^{2\pi/\Omega} dt\ \cos n\Omega t \cdot E_\rho(\rho, d = \cos \Omega t) \qquad \text{(Eq 16)}$$

Since the lightning rod model predicts a physically meaningful electric field profile for all probe-sample distances $d$, the power distribution $P_n(\rho)$ for $n = 2{,}3$ was straightforwardly calculated with the relevant products of demodulated field distributions $E_{\rho,n}(\rho)$ inclusive of reflected fields from the sample.

Supplementary Figure 8 shows the field and temperature profiles for several frequencies. We note that the $dT/dx$ profile is qualitatively similar to our observed photocurrent pattern. Let's say, the Seebeck profile is narrow compared to the $dT/dx$ such that it can be approximated as a delta function. Then, from Eq 5, we see that the photocurrent profile will be identical to $dT/dx$. Therefore, we conclude that any Seebeck coefficient profile that is significantly narrower than the cooling length will produce a photocurrent pattern that is consistent with our experimental data.

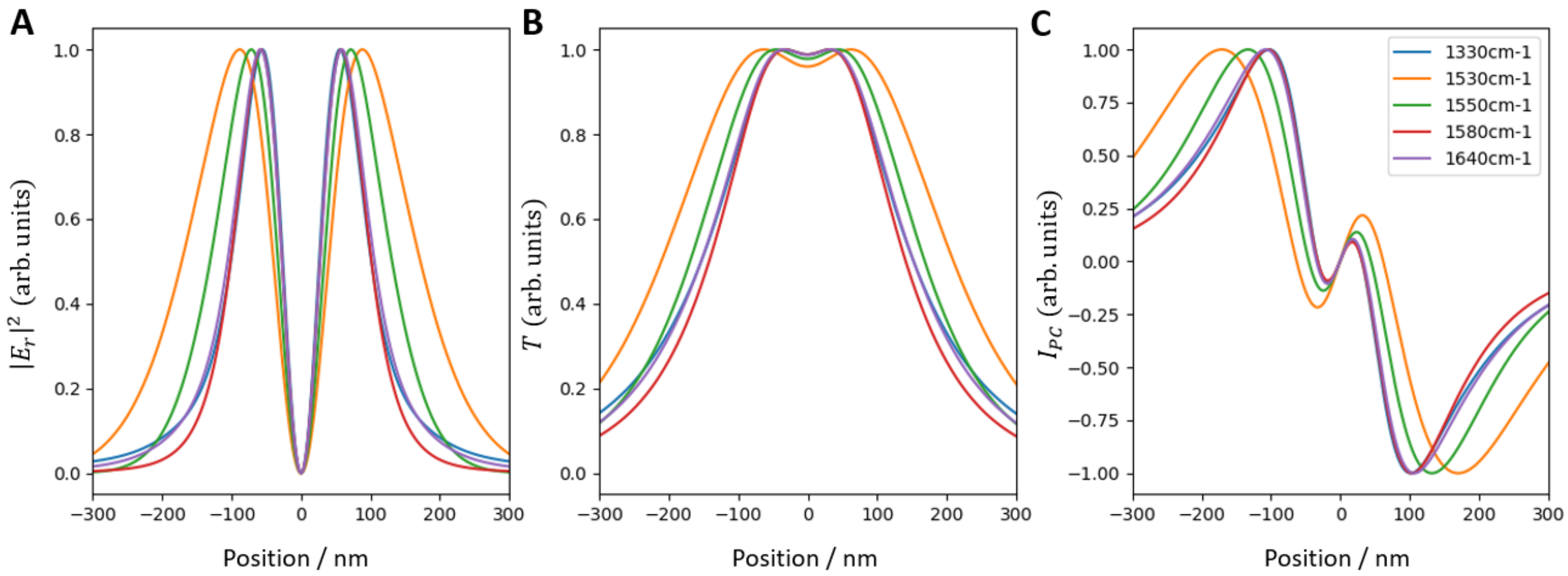

**Supplementary Figure 8 | Electric field and temperature profiles.** (A - C) Radial electric field $E_r$, hot carrier temperature $T$ and $dT/dx = \hat{x} \cdot \nabla T$profiles at various frequencies. The tip is located at the origin.

## Supplementary Note 3.5: Converting 1D profiles to 2D profiles – superposition model

To convert the 1D profiles calculated in Supplementary Note 3.2 into 2D profiles, we used a simple superposition model. However, the superposition model may not accurately reproduce the Seebeck profile at the AA sites. Here, we compare the relative importance of the domain walls and the AA sites to the calculated photocurrent pattern by separating their relative contributions.

First, we define a mask which is a series of Gaussians centered on the AA sites. Let the $n$ AA sites be located at $\{x_n, y_n\}$. Then, the mask is given by

$$M(x,y) = \sum_n \exp\left(-\frac{(x-x_n)^2 + (y-y_n)^2}{w_{AA}^2}\right) \quad \text{(Eq 35)}$$

where $w_{AA}$ is the width of the Gaussians. Then we separate the Seebeck coefficient at the AA sites by multiplying the Seebeck coefficient from the superposition model by the mask:

$$S_{AA}(x,y) = S_{2D}(x,y)M(x,y) \quad \text{(Eq 36)}$$

The domain wall contribution is then

$$S_{DW}(x,y) = S_{2D}(x,y)(1 - M(x,y)) \quad \text{(Eq 37)}$$

such that

$$S_{AA}(x,y) + S_{DW}(x,y) = S_{2D}(x,y) \quad \text{(Eq 38)}$$

Furthermore, since convolution is linear, the following is also true:

$$I_{PC,AA} + I_{PC,DW} = I_{PC} \quad \text{(Eq 39)}$$

where $I_{PC,AA}$, $I_{PC,DW}$ and $I_{PC}$ are the photocurrent patterns arising from $S_{AA}$, $S_{DW}$ and $S_{2D}$ respectively.

Supplementary Figure 16 shows the Seebeck coefficient and photocurrent patterns arising from the profiles calculated above. We see that $I_{PC,AA}$ is simply a series of dipoles centered at the AA sites and does not resemble the pattern observed in the experiment. At the same time, $I_{PC,DW}$ reproduces both the meandering pattern as well as the fine features at the domain walls. The spatial patterns in the sum $I_{PC}$ are only slight modifications to $I_{PC,DW}$. Therefore, we conclude that the 1D Seebeck coefficient variation across the domain wall is dominant in explaining the observed experimental pattern, thus justifying the use of the superposition model.

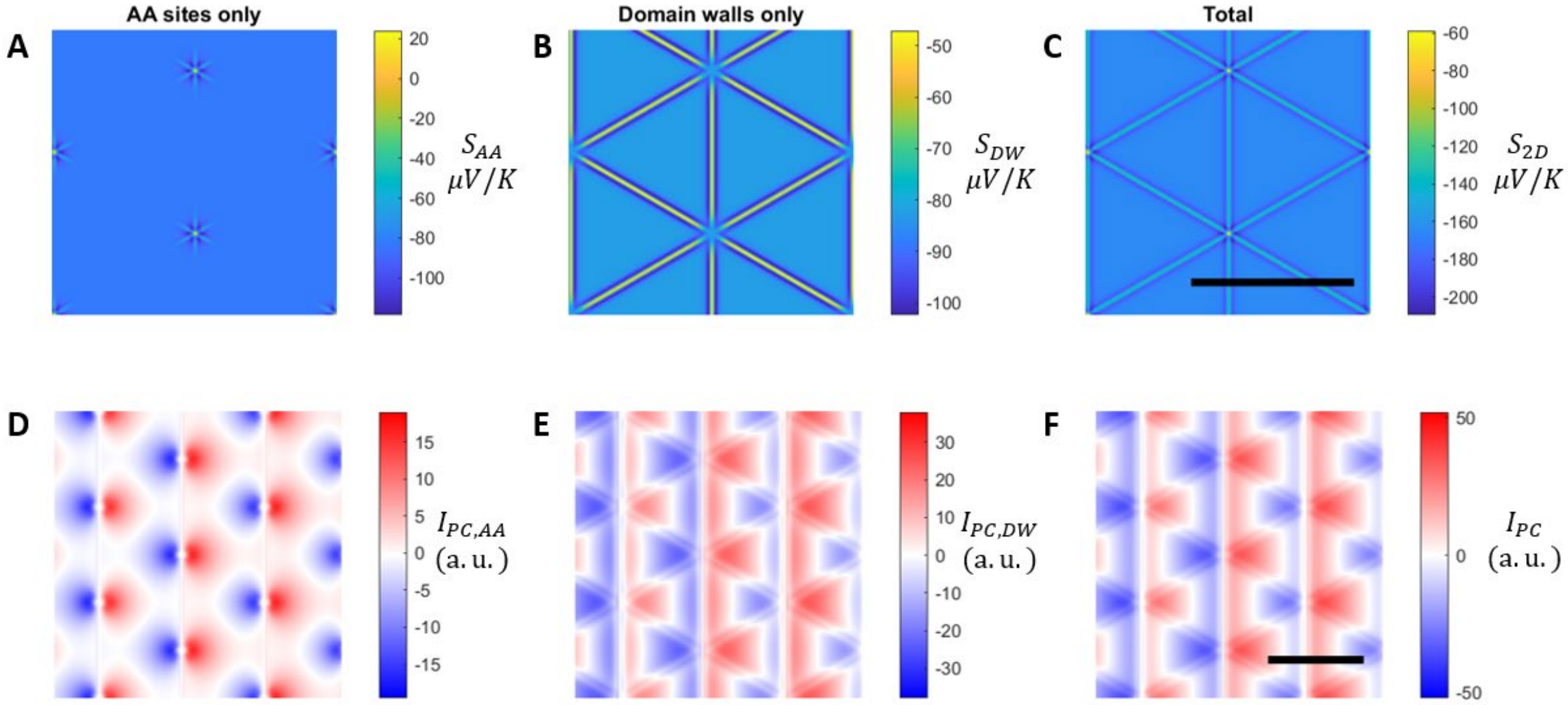

**Supplementary Figure 16 | Relative importance of the AA sites and the domain walls to the calculated photocurrent pattern.** (A) Seebeck coefficient of the AA sites only $S_{AA}$ (B) Seebeck coefficient of the domain walls only $S_{DW}$ (C) Total Seebeck coefficient calculated with the superposition model $S_{2D}$. (D – F) Calculated photocurrent patterns for the Seebeck coefficients in (A – C). Scale bars 500 nm.

## Supplementary References